\documentclass[12pt]{article}
\usepackage{epsfig}
\usepackage[hang,bf,small]{caption}
\DeclareGraphicsExtensions{.eps.gz,.eps,.ps,.ps.gz}
\parskip=\itemsep               

\newcommand{\beq}{\begin{equation}}
\newcommand{\eeq}{\end{equation}}
\newcommand{\beqar}[1]{\begin{eqnarray}\label{#1}}
\newcommand{\eeqar}{\end{eqnarray}}

\newcommand{\si}{\sigma}

\newcommand{\as}{\alpha_S}

\def\eq#1{{Eq.~(\ref{#1})}}

\def\arnps#1#2#3{  {\it Ann. Rev. Nucl. Part. Sci. }{\bf #1} (19#2) #3}
\def\npb#1#2#3{    {\it Nucl. Phys. }{\bf B#1} (19#2) #3}

\def\prd#1#2#3{    {\it Phys. Rev. }{\bf D#1} (19#2) #3}

\def\zpc#1#2#3{    {\it Z. Phys. }{\bf C#1} (19#2) #3}

\relax
\begin{document}
\title{
{\Large \bf   Parton Densities and Saturation Scale }\\\
{ \Large \bf  from Non-Linear Evolution  in  DIS  on Nuclei}}
\author{
{\large  ~ E.~Levin\thanks{e-mail: leving@post.tau.ac.il}~~$\mathbf{{}^{a)}}$ \,~\,and\,\,~
M. ~Lublinsky\thanks{e-mail: mal@techunix.technion.ac.il}~~$\mathbf{{}^{b)}}$}\\[4.5ex]
{\it ${}^{a)}$ HEP Department}\\
{\it  School of Physics and Astronomy}\\
{\it Raymond and Beverly Sackler Faculty of Exact Science}\\
{\it Tel Aviv University, Tel Aviv, 69978, ISRAEL}\\[4.5ex]
{\it ${}^{b)}$  Department of Physics}\\
{\it  Technion -- Israel Institute of   Technology}\\
{\it  Haifa 32000, ISRAEL}\\[1.5ex]
}

\maketitle
\thispagestyle{empty}
                      
\begin{abstract} 
~\,\, We present the numerical solution of the non-linear evolution equation
for DIS on nuclei  for $x = 10^{-2} \div 10^{-7}$. We demonstrate that the
solution to the non-linear evolution equation is quite different from the Glauber
- Mueller formula which was  used as the initial condition for the
equation.

We illustrate the energy profit for performing  DIS experiments on nuclei.
 However, it
turns out that the gain is quite modest: $x_{Au} \simeq  5\, x_{\rm proton} $ 
for the
same parton density.

 We find that the  saturation
scale $Q^2_s \propto A^{\frac{1}{3}}$. For gold  the saturation scale 
  $Q_{s,Au} \simeq 1.5\,\, GeV$  at  $x= 10^{-3}$. 
Such a  large value  leads to considerable
contribution of the high density QCD phase to RHIC data and reveals itself in 
essential damping for both   $xG_A$ and $F_{2A}$.

 \end{abstract}
\thispagestyle{empty}
\begin{flushright}
\vspace{-19.5cm}
TAUP-2670-20001\\
\today
\end{flushright}   
\newpage
\setcounter{page}{1}

\section{Introduction}
\setcounter{equation}{0}

Deep inelastic scattering on nuclei gives us a new possibility to reach a  high
density QCD phase without requiring extremely low values of $x$ (extremely
high  energies). This can be seen directly from  simple estimates
of the packing factors for the partons in DIS on nuclei. Indeed, the packing
factor is equal to
\beqar{KAPPA}
\kappa_A(x,Q^2)\,\,&=&\,\,\frac{3 \as \,\pi^2}{Q^2}\times \frac{1}{\pi R^2_A}\,\times\,
xG^{DGLAP}_A(x,Q^2) \nonumber\\
                   &=&   \frac{3 \as \,\pi^2}{Q^2}\times \frac{1}{\pi R^2_A}
                     \,\times\,A\,xG^{DGLAP}_N(x,Q^2) \nonumber\\
               &\approx & A^{\frac{1}{3}}\,\kappa_N(x,Q^2) 
\eeqar
For  a gold nucleus  the packing factor is the same as for a nucleon
at  $x_A \approx 200 \cdot x_N$, assuming  $xG_N(x,Q^2) \propto (1/x)^{\lambda}$
with $\lambda \sim 0.3 $ which follows from the HERA experimental data \cite{HERAREV}.
The gluon density $xG^{DGLAP}$ is the solution of the linear DGLAP evolution
equation \cite{DGLAP}.  The  equation (\ref{KAPPA}) implies another advantage of 
nuclear targets.  Namely, the value of the saturation scale is larger for  nuclear 
targets than for  nucleon ones. Indeed, the simplest estimate for the saturation
scale can be deduced  from the equation $ \kappa_A(x,Q^2_{s,A}(x))
\,=\,1$. The solution of this equation is 
\beq \label{SATSCEST}
Q^2_{s,A}(x) \,\,\propto\,\,A^{\frac{1}{3 ( 1\,-\,\gamma)}}\,\,,
\eeq
where the anomalous dimension $\gamma= d \ln (x G_N)/ d \ln Q^2$ is a smooth
function of $x$ and $Q^2$. The fact that the saturation scale is larger for nuclei 
from the theoretical point of view makes our calculations more reliable. 

Determination of the saturation scale $Q_{s,A}(x)$ is one  of the most challenging 
problems in QCD.
The DGLAP equation \cite{DGLAP} describes the gluon radiation which increases the number
of partons. However, when the parton density becomes large the annihilation
processes enter  the game and they  suppress the gluon radiation.  These effects
 tame the rapid increase of the parton densities at the saturation
scale $Q_{s,A}(x)$  \cite{GLR,MUQI,MV}.  
At this scale  the nonlinear effects become important and the parton evolution 
cannot be described by a linear equation any more. This argument 
stimulated  a development of new methods 
\cite{GLR,MUQI,MV,SAT,AGL,MUSAT,KM,ELTHEORY,BA,KO,ILM,IM,Braun}, which
finally lead to  the very same nonlinear evolution equation for the imaginary part of the
DIS elastic amplitude\footnote{
\eq{EQ} was  
proposed  in momentum representation by Gribov, Levin and Ryskin \cite{GLR} and it
was proved in double log approximation of perturbative QCD by Mueller and Qiu
\cite{MUQI}, in
Wilson Loop Operator Expansion at high energies by Balitsky \cite{BA}, in
colour dipole approach \cite{MU94} to high energy scattering in QCD  by
Kovchegov \cite{KO}, 
 in effective
Lagrangian approach for high parton density QCD by Iancu, Leonidov and McLerran 
(see Ref. \cite{ILM} and
Refs. \cite{ELTHEORY} for previous efforts). Braun \cite{Braun} derived this
equation by summing ``fan'' diagrams, as GLR did, but using the triple BFKL ladder
vertex of Refs. \cite{3P}. 
 Therefore, this equation is a reliable
tool for an extrapolation of the parton distributions to the region of low $x$.  
}:
\begin{eqnarray}
\label{EQ} 
 & &   N_A({\mathbf{x_{01}}},Y;b)  \,=\,  N_A({\mathbf{x_{01}}},Y_0;b)\, 
{\rm exp}\left[-\frac{2
\,C_F\,\as}{\pi} \,\ln\left( \frac{{\mathbf{x^2_{01}}}}{\rho^2}\right)(Y-Y_0)\right ]\,
+\nonumber  \\ & & \frac{C_F\,\as}{\pi^2}\,\int_{Y_0}^Y dy \,  {\rm exp}\left[-\frac{2
\,C_F\,\as}{\pi} \,\ln\left( \frac{{\mathbf{x^2_{01}}}}{\rho^2}\right)(Y-y)\right ]\,\times
\\ & &\int_{\rho} \, d^2 {\mathbf{x_{2}}}  
\frac{{\mathbf{x^2_{01}}}}{{\mathbf{x^2_{02}}}\,
{\mathbf{x^2_{12}}}} \nonumber 
\left(\,2\,  N_A({\mathbf{x_{02}}},y;{ \mathbf{ b-
\frac{1}{2}
x_{12}}})-  N_A({\mathbf{x_{02}}},y;{ \mathbf{ b -
\frac{1}{2}
x_{12}}})  N_A({\mathbf{x_{12}}},y;{ \mathbf{ b- \frac{1}{2}
x_{02}}})\right)\,. \nonumber
\end{eqnarray}

The equation is written for $N_A(r_{\perp},x; b) = Im
\,a^{el}_{\rm dipole}(r_{\perp},x; b)$, where $a^{el}_{\rm dipole}$ is the amplitude
of the elastic scattering for the dipole of the size $r_{\perp}$. The total
dipole cross section  is given by
\beq \label{TOTCX}
\sigma^A_{\rm dipole}(r_{\perp},x) \,\,=\,\,2\,\,\int\,d^2
b\,\,N_A(r_{\perp},x;b)\,.
\eeq
The total deep inelastic cross section  is related to the dipole
cross section
\beq
\label{F2}
\si_A(x,Q^2)\,\,\,=\,\,\int\,\,d^2 r_{\perp} \int \,d
z\,\,
|\Psi^{\gamma^*}(Q^2; r_{\perp}, z)|^2 \,\,\sigma^A_{\rm dipole}(r_{\perp},
x)\,,
\eeq 
where the QED wave functions $\Psi^{\gamma^*}$ of the virtual photon are well known
\cite{MU94,DOF3,WF}. The equation (\ref{F2}) is quite transparent and it  describes   two
stages of  DIS \cite{GRIB}. The  first one is the  decay of a virtual photon into colourless
dipole ($ q \bar q $ -pair) which is described by the wave function
$\Psi^{\gamma^*}$ in \eq{F2}.
The second stage is the interaction of the dipole
with the target ($\sigma^A_{\rm dipole}$ in \eq{F2}). This equation is the
simplest manifestation of the fact that the correct degrees of freedom at
high energies in QCD are colour dipoles \cite{MU94}.

In the equation (\ref{EQ}), 
the rapidity $Y=-\ln x$ and $Y_0=-\ln x_0$. The ultraviolet cutoff $\rho$
is needed to regularize the integral, but it does not appear in physical quantities. In the large
 $N_c$ (number of colours) limit  $C_F=N_c/2$.
\eq{EQ} has a very simple meaning:  the dipole of size $\mathbf{x_{10}}$ decays
into two dipoles of  sizes $\mathbf{x_{12}}$ and $\mathbf{x_{02}}$  with the decay probability
given by  the wave
function  $| \Psi|^2
\,=\,\frac{\mathbf{x^2_{01}}}{\mathbf{x^2_{02}}\,\mathbf{x^2_{12}}}$. 
 These two dipoles
 then interact with the target. The non-linear term  takes into account
the Glauber corrections for such an interaction.

The equation (\ref{EQ}) does not depend on the  target explicitly.
This independence is a direct indication that the equation is
  correct for any target  in the regime of high parton density. The only dependence on a target
comes from  initial conditions at some initial value $x_0$.
For a nucleus target it was proven  in Refs. \cite{AGL,KO}   that the initial conditions  should be
  taken in the Glauber form:
\beq
\label{ini}
  N_A(\mathbf{x_{01}},x_0;b)\,=\,N_{GM}^A(\mathbf{x_{01}},x_0;b)\,,
\eeq
with 
\beq
\label{Glauber}
N_{GM}^A(\mathbf{x_{01}},x;b)\,=\,1\,\,-\,{\rm exp}\left[ - 
\frac{\sigma^N_{\rm dipole}(r_{\perp},x)}{2}
\,\,S_A(b)\right]\,.
\eeq
The cross section of the dipole interaction with a nucleon is defined
as 
 $$\sigma^N_{\rm dipole}(r_{\perp},x) \,\,=\,\,2\,\,\int\,\,d^2 \,b
\,\,N_N(r_{\perp},x;b)$$
with $N_N$ that has been calculated in the  Ref. \cite{LGLM}.
The equation (\ref{Glauber}) represents the Glauber - Mueller  (GM) formula 
which accounts for the  multiple
dipole-target interaction in the eikonal approximation \cite{DOF3,DOF1,DOF2}.

It turns out, however,  that at sufficiently high initial $x_0$ ($x_0 \,\approx\,10^{-2}$) 
the equation (\ref{Glauber}) leads to
 the same initial profile $N_A$ as the  simplified version
\beq \label{GLAUBSI}
N_{GM}^A(\mathbf{x_{01}},x;b)\,=\,1\,\,-\,{\rm exp}\left[ - 
\frac{\as \pi^2  \mathbf{x^2_{01}}}{2\,N_c}
\,xG^{DGLAP}(x,  4/\mathbf{x_{01}^2})\,S_A(b)\right]\,\,.
\eeq

The nucleon gluon distribution inside a  nucleus $xG^{DGLAP}$ is parameterized according
to LO solution of the DGLAP equation \cite{DGLAP}.  
The function $S_A(b)$ is a  dipole  profile function inside the nucleus  with the  atomic number 
$A$.  The value  of $x_0$ is chosen within the interval 
\beq \label{INCON}
{\rm exp}( - \frac{1}{\as} ) \,\leq \,x_0 \,
\leq \frac{1}{2 m R_A}\,\,,
\eeq
where $R_A$ is the radius of the target. In this region the value of $x_0$ is small 
enough to use the low $x$ approximation but the production of the gluons 
(colour dipoles) is still suppressed since $\alpha_S \ln (1/x) 
\,\leq\,\,1$. 
Therefore in this region we have instantaneous exchange of the classical gluon fields.    
Due to  this fact an incoming 
colour dipole interacts separately with each  nucleon in a nucleus \cite{KM}. 

Solutions to the equation (\ref{EQ}) were studied in asymptotic limits in the 
Refs. \cite{MUSAT,IM,LT,LT1}.
A first  attempt to estimate numerically the value of the shadowing corrections
in DIS on nuclei in the framework of the non-linear evolution equation was
made in the Ref. \cite{AGL}. In that paper  the AGL evolution equation was solved
in the semi-classical approximation. It turns out that the AGL equation is the
same as \eq{EQ} in the double log approximation but it is written not for the
amplitude $N$ but for the opacity $\Omega$ ($N = 1 - \exp(- \Omega/2)$).
Braun \cite{Braun}  solved numerically an equation similar to 
\eq{EQ} in momentum representation  
with some specific initial conditions and we will discuss his results below. In our recent paper 
\cite{LGLM} we solved \eq{EQ} numerically for the proton targets.

Our  approach to  the solution of \eq{EQ} is based on space representation. We have several
arguments for working in this representation. First of all, 
 it allows  us to  formulate the initial conditions. Moreover,  the initial conditions are not
known at large distances at all. Consequently  we would expect problems calculating 
the amplitude in the momentum 
representation  since it implies  knowledge 
of the amplitude in the whole kinematic region of $r_{\perp}$.
  Our last argument is that  in the space representation 
the unitarity constraints  have  the  simplest form:
\beq \label{UNTY}
2\,N(r_{\perp},x;b)\,\,=\,\,N^2(r_{\perp},x;b)\,\,+\,\,G_{in}(r_{\perp},x;b)\,\,,
\eeq
where $G_{in}$ stands for  contributions of all inelastic processes. We assume in
\eq{UNTY} that at high energies the elastic amplitude  is dominantly imaginary
 $ a^{el}(r_{\perp},x;b)\,\,\longrightarrow\,\,N(r_{\perp},x;b) $. The equation
(\ref{UNTY}) implies $N(x,r_{\perp};b) \,\leq\,1$ with a natural limit
$N(x,r_{\perp};b)\,\rightarrow\,1$ at $x \,\rightarrow\,0$. This limit provides us with a
very useful check in our numerical approach.

In the present paper we report on our solution of the equation (\ref{EQ}) for the nuclear 
targets. The method of iterations proposed in  \cite{LGLM} is applied. 
Contrary to the Ref. \cite{Braun}
we study solutions for phenomenologically reasonable kinematic regions only, and for real
nuclei. The main goal  of the research
is determination of the saturation scale $Q_{s,A}(x)$ and its dependence on the target 
atomic number  $A$. 
We also present  estimates of the gluon density  $xG_A$ and the structure function $F_{2A}$
in the region of the high parton density QCD.
 
The paper is organized as follows. In the next section (2) we give a review of the nuclear target
properties. In section 3 a solution of the nonlinear equation is presented. Section 4
is devoted to the calculations of $xG_A$ and $F_{2 A}$.  In  section 5 we
determine  the saturation scales and study their $A$-dependence. 
Section 6 presents  discussion of 
 the self-consistency of the numerical procedure and results. 
  We conclude in the last section (7).

\section{Nuclear  input data}

In this section we discuss  the properties of  the nuclei our computations are made for. 
The nuclei are 
described by their radial  nucleon distribution  $\rho(r)$. 
This distribution is normalized to the total nucleon number $A$:
\beq
\label{rho}
\int\rho(r)\,d^3r\,=\,A\,.
\eeq
The nucleon distribution for all nuclei discussed in this paper are parameterized by one of the 
following three parameterizations:
\begin{enumerate}
\item 
Two-parameter Fermi model (2pF)
$$
\rho(r)\,=\,\rho_0/(1\,+\,exp\,((r-c)/z))\,.
$$
\item
Three-parameter Fermi model (3pF)
$$
\rho(r)\,=\,\rho_0\,(1\,+\,w\,r^2/c^2)/(1\,+\,exp\,((r-c)/z))\,.
$$
\item
Three-parameter Gaussian model (3pG)
$$
\rho(r)\,=\,\rho_0\,(1\,+\,w\,r^2/c^2)/(1\,+\,exp\,((r^2-c^2)/z^2))\,.
$$
\end{enumerate}
The constants $c$, $z$, and $w$ are free parameters varying from nucleus to nucleus. 
The table  (\ref{t1}) lists the relevant input data for all  nuclei used in this work \cite{NDATA}.
           The parameter $\rho_0$ is always found due to the normalization (\ref{rho}).
\begin{table}
\center{
\begin{tabular}{||l||c|c|c|c|c|c||} 
\hline
    Nucleus                     & $A$ &  model &$ R_A$ & $c$ &  $z$ & $w$     \\
                         &        &              &         [fm]                          &  [fm] &  [fm]&              \\
\hline \hline
 $Ne$ &                    20   &  2pF      &    3                       &  2.805 & 0.571    &    \\  
\hline
$Ca$ &                     40 &   3pF        & 3.486                    & 3.6685   &  0.5839  & -0.1017    \\
\hline
 $Zn$ &                    70 &   2pF       &   4.044                 &   4.409   &   0.583 &    \\
\hline
 $Mo$ &                    100 &   3pG  &     4.461                &   4.559   &    2.6723 &  0.339    \\
\hline
 $ Nd$ &                    150 &   2pF  &     5.048                &   5.7185   &    0.651 &      \\
\hline
$Au$  &                    197 &  2pF  &      5.33                  &    6.38    &    0.535   &  \\
\hline
\end{tabular}}
\caption{Parameters for the nucleon distributions in the nuclei taken from the Ref. \cite{NDATA}.}
\label{t1}
\end{table}

\begin{figure}[htbp]
\epsfig{file=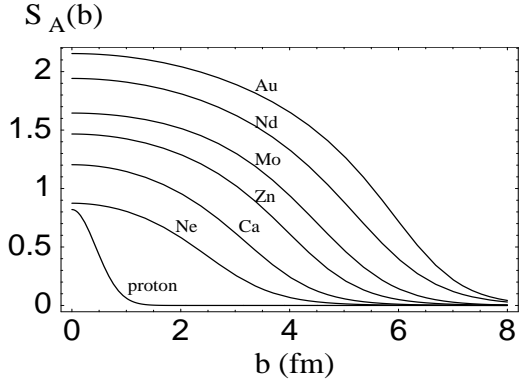,width=70mm, height=50mm}
\begin{minipage}{10.0 cm}
\vspace{-6cm}
\caption[]{The nucleon transverse profile function inside the nucleus $S_A(b)$  
is plotted as a function of  $b$.}
\label{profile}
\end{minipage}
\end{figure}

The  radius $r$ can be decomposed to the transverse $b$ and longitudinal $r_l$ 
components
$$
r^2\,=\,b^2\,+\,r_l^2.
$$
Then the transverse profile function $S_A(b)$ is defined 
as the integral of the radial density  $\rho$ over the longitudinal  component $r_l$:
\beq
\label{Sb}
S_A(b)\,=\,\int \rho(r) \,dr_l.
\eeq
Fig.~\ref{profile} displays the function  $S_A(b)$ for various nuclei.

We wish to draw reader's attention to the fact that $S(b = 0)$ for  proton is
almost the same as for $Ne$ (see Fig.~\ref{profile}). 
This fact has an obvious explanation:  partons
inside  the proton are distributed in smaller area with $R \approx
3\,GeV^{-1}\,\,=\,\,0.6\,fm$  which is smaller than the area given by
extrapolation of the Wood-Saxon parameterization to $A=1$. Indeed, 
 $R_{A=1} \,\simeq\,\,1.2\,fm \,\,>\,\,R=0.6\,fm$. This simple fact makes our
 estimates  based on \eq{KAPPA}  too optimistic. We will show below that the
 packing factor for gold is the same as for  nucleon at $x_{Au} \,\approx\,(\,5
 \,\div\,10 \,) \cdot x_N$ .

\section{Solution of the nonlinear equation}

In the Ref. \cite{LGLM} we suggested to solve the equation (\ref{EQ}) by the method 
of iterations. All details about our method can be found in that paper, where the equation
(\ref{EQ}) was solved numerically for the proton target. 

Following the very same procedure as in  \cite{LGLM}  we obtain numerical solutions 
for all nuclei listed above. For $xG^{DGLAP}(x,Q^2)$  we use 
the  GRV'94 parameterization and  
the leading order solution of the DGLAP evolution equation \cite{GRV}.
The initial conditions (\ref{ini}) are set at $x_0=10^{-2}$. 
The constant value for the strong coupling constant $\as=0.25$ is always used. The solutions
are computed  within the kinematic region down to $x=10^{-7}$ and distances
up to a few fermi. 

The function $N_A$ is formally a function of three variables:
 the energy variable $x$, the transverse
distance $r_\perp$, and the impact parameter $b$. The $b$-dependence is  parametric only 
because the  evolution kernel does not depend on $b$. In order to simplify the problem 
we assume that the function $N_A$ preserves the very same $b$-dependence as introduced 
in the initial conditions:\footnote{Note 
that for the Gaussian parameterization of the profile function $S_A$, 
$A=\pi \,R_A^2\,S_A(0)$.}
\beq
\label{Nb}
 N_A(r_\perp,x; b)\,=\, (1\,-\,e^{-\kappa_A(x,r_\perp)\, S_A(b)/S_A(0)}).
\eeq
The function $\kappa_A$ is related to the ``$b=0$'' solution $\tilde N_A(r_\perp,x)$:
\beq
\label{kappa}
\kappa_A(x,r_\perp)\,=\,-\,\ln(1\,-\,\tilde N_A(r_\perp,x)).
\eeq
$\tilde N_A(r_\perp,x)$ represents a solution of the very same equation (\ref{EQ}) but with
no dependence on the third variable. The initial conditions for the function 
$\tilde N_A(r_\perp,x)$ are taken at $b=0$. For the case of the proton target \cite{LGLM} 
the anzatz in the form (\ref{Nb}) is shown to be a quite  good approximation of the exact
$b$-dependence of the solution to (\ref{EQ}). We will investigate the accuracy of this 
anzatz for gold  in the end of this section.

\begin{figure}[htbp]
\begin{tabular}{c c}
 \epsfig{file=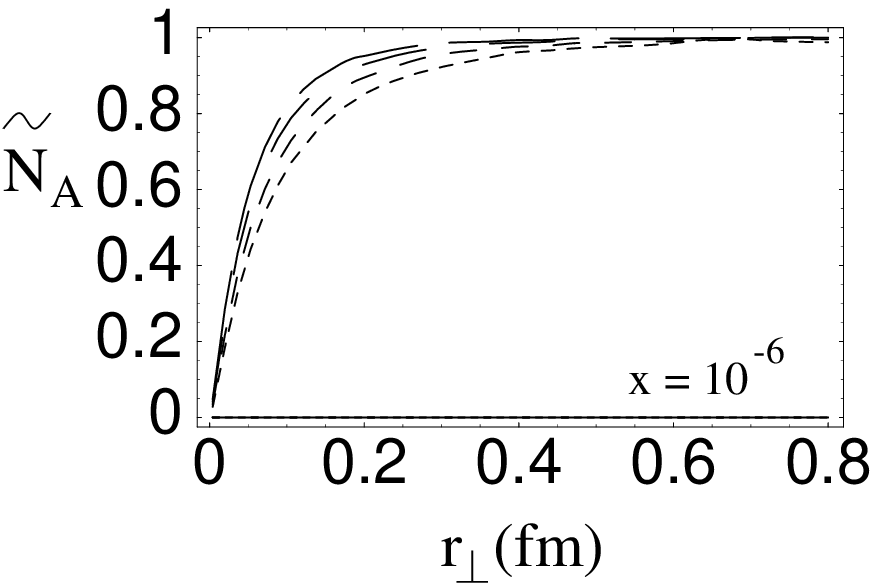,width=70mm, height=50mm}&
\epsfig{file=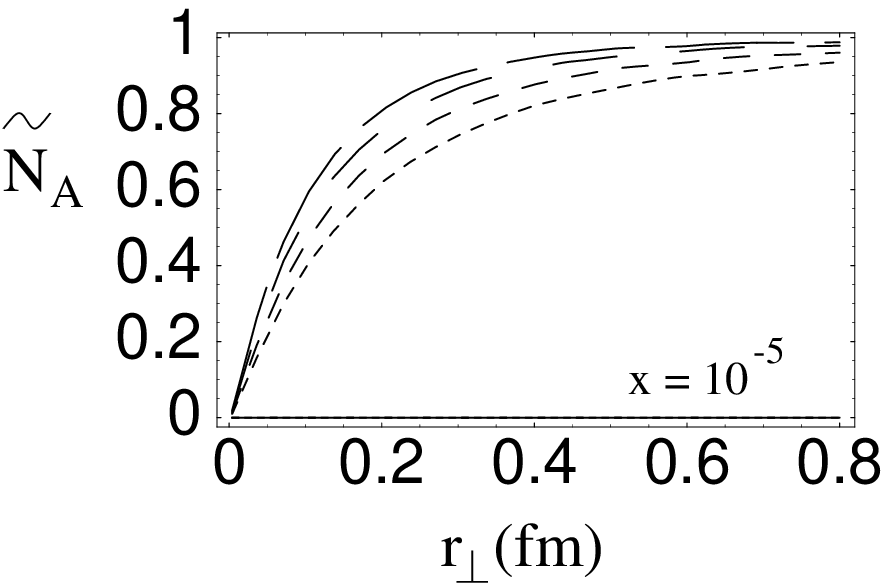,width=70mm, height=50mm}\\ 
 \epsfig{file=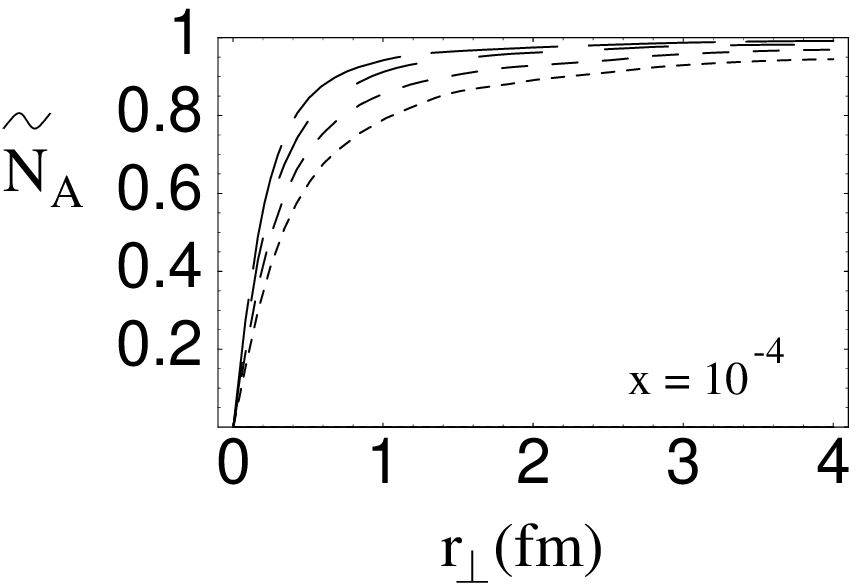,width=70mm, height=50mm}&
\epsfig{file=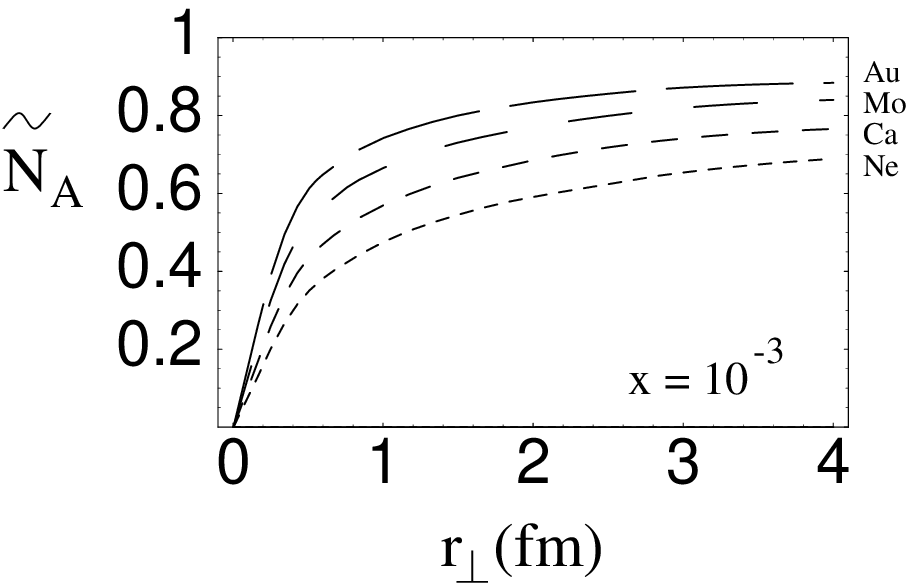,width=73mm, height=50mm}\\ 
\end{tabular}
  \caption[]{\it The function $\tilde N_A$ is plotted versus distance. The four curves
 show the result for the nuclei  $Ne$, $Ca$, $Mo$, and $Au$. }
\label{solution}
\end{figure}

For each nucleus  the function $\tilde N_A(r_\perp,x)$ is obtained after about ten iterations. 
The Fig.~\ref{solution} shows the solutions for various nuclei. Note that though the solutions 
are of similar form they differ from each other indicating some memory of the initial conditions.

\begin{figure}[htbp]
\begin{tabular}{c c}
 \epsfig{file=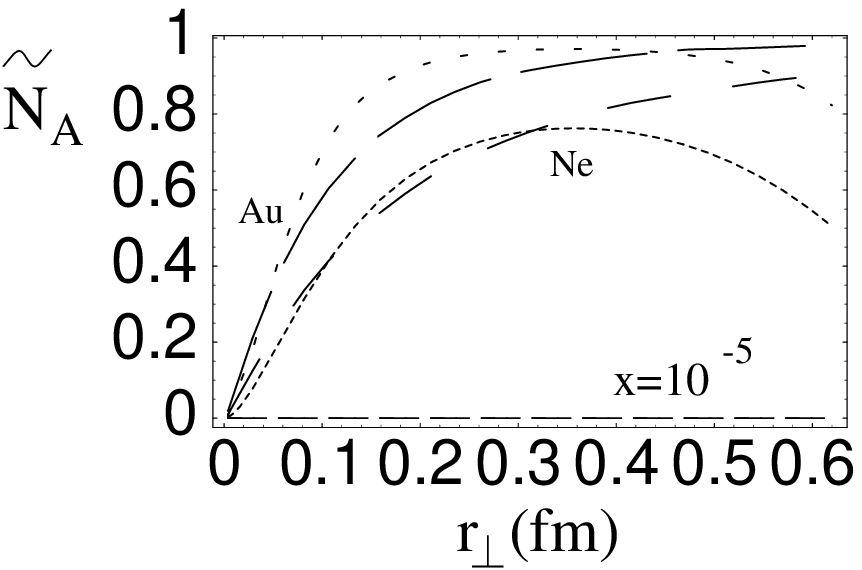,width=70mm, height=50mm}&
\epsfig{file=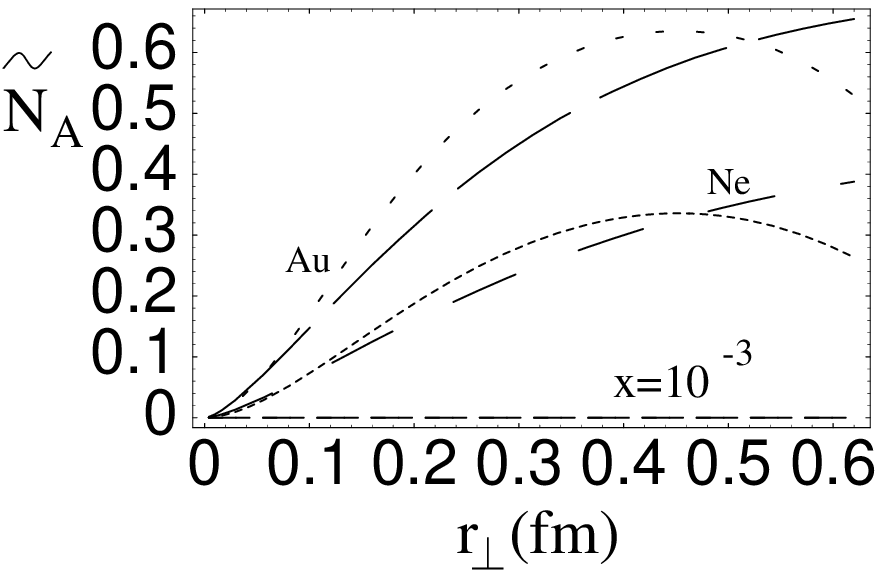,width=70mm, height=50mm}\\ 
\end{tabular}
  \caption[]{\it The functions $\tilde N_A$ for $Au$ and $Ne$ (dashed lines) 
are  plotted together with the corresponding 
Glauber formulas (dotted lines). }
\label{comp}
\end{figure}

It is  worth to compare the obtained solutions with the Glauber formula (\ref{GLAUBSI}).
For example the comparison is made for two nuclei: 
the lightest one $Ne$ and the most heavy $Au$. The two
models are plotted together in the  figure \ref{comp}.  

Having obtained the solutions for nuclei we can answer the question about the energetic 
gain of performing DIS on nuclei.  To this goal the results for nuclei should be compared with
ones for  proton targets \cite{LGLM}. Unfortunately,  the gain is quite  modest and it consists
of about half order of magnitude for very  heavy nuclei. Namely, for a gold target $Au$ the 
scattering amplitude $\tilde N_{Au}$ (and consequently the packing factor $\kappa_{Au}$)
 almost coincides with the proton one 
when $x_{Au}\simeq 5\, x_{\rm proton}$, which is quite different from the previous optimistic
estimates (see \eq{KAPPA} and discussion there). The simple reason for
such a mild gain was mentioned in the previous section and it is related to
the fact that $S_N(b = 0)$ is almost the same as for $Ne$ as can be seen in the Fig.~\ref{profile}.

\begin{figure}[htbp]
\begin{tabular}{c c}
\epsfig{file=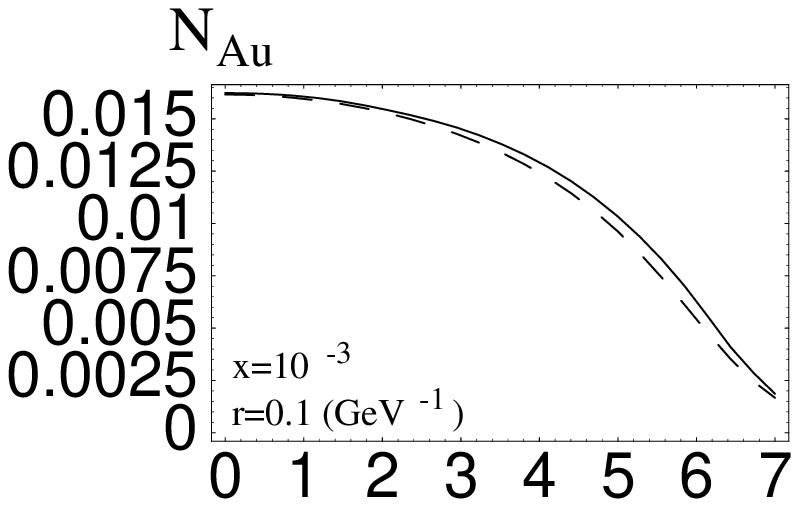,width=77mm, height=50mm}&
\epsfig{file=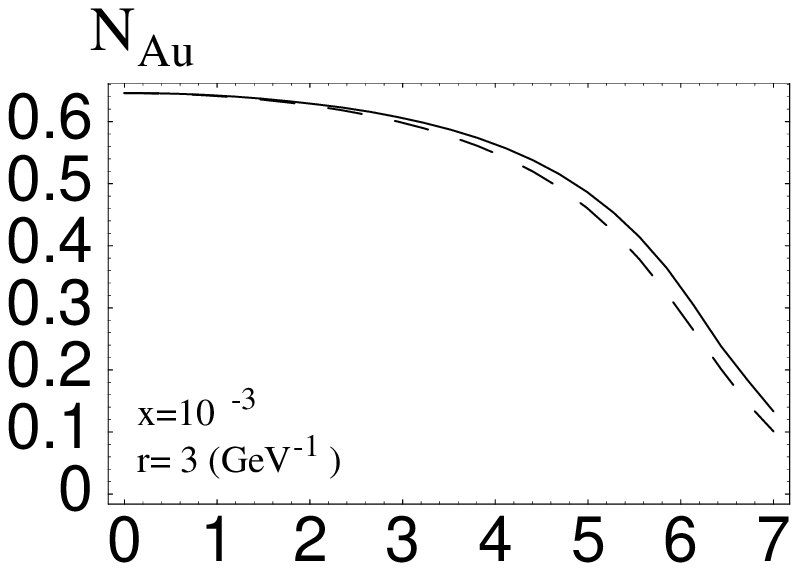,width=70mm, height=50mm}\\ 
\hspace{0.6cm} \epsfig{file=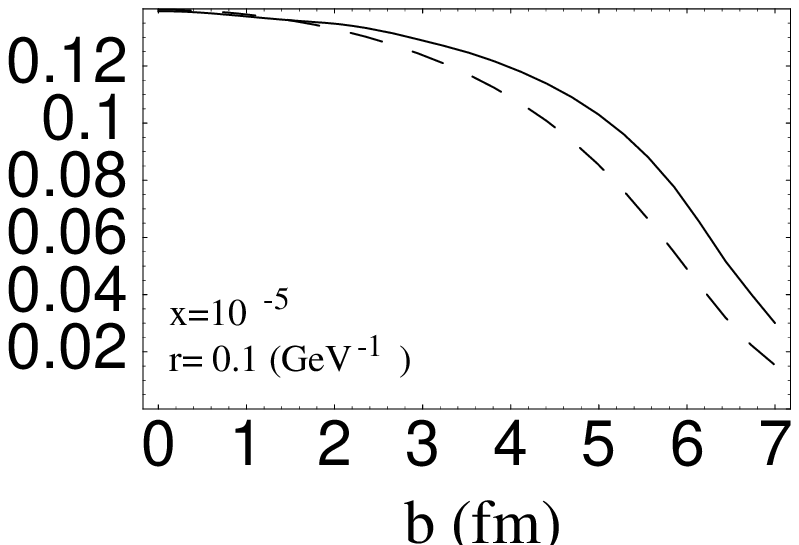,width=70mm, height=50mm}&
\epsfig{file=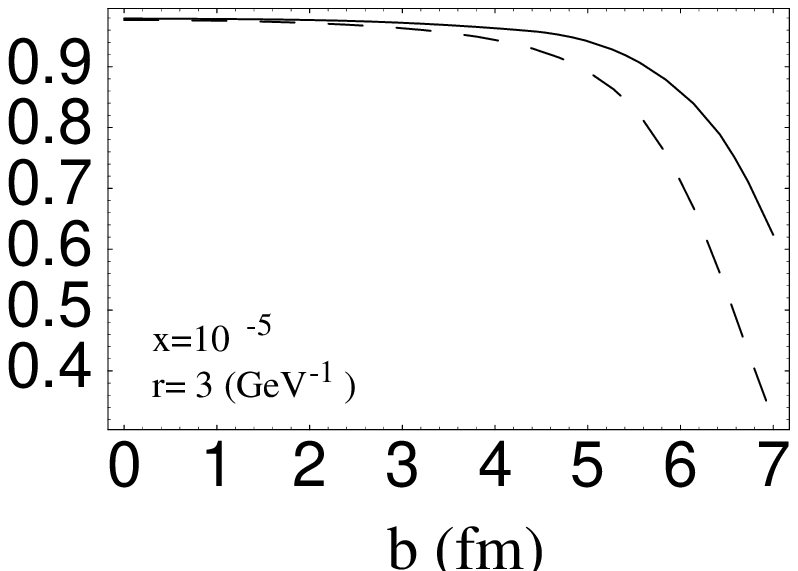,width=70mm, height=50mm}\\ 
\end{tabular}
  \caption[]{\it The function $N_{Au}$ is plotted versus the impact parameter $b$. 
The solid line is the exact $b$ dependence, while the dashed line is the anzatz (\ref{Nb}). }
\label{bdep}
\end{figure}

It is possible to determine the gluon density $xG_A$ from the solutions obtained. To this goal
  the accuracy of the anzatz (\ref{Nb}) has to be estimated. 
The $b$-dependence so far ignored in the function $\tilde N_A$ 
should be investigated.  In order to restore the $b$-dependence of the function $N_A$
we solve the equation (\ref{EQ}) with the only assumption that the impact parameter 
$b$ is much larger than the dipole sizes:
\beq
\label{assum}
{\mathbf{x_{01}}}\,\ll\,b; \,\,\,\,\,\,\,\,\,\,\,\,\,\,\,\,\,\,{\mathbf{x_{02}}}\,\ll\,b.
\eeq
In this case the $b$-dependence of the rhs of (\ref{EQ}) is significantly simplified. We perform
the computations for only one nucleus $Au$, 
which is the most different from the proton among the nuclei that we have
 considered.
The Fig. \ref{bdep} shows examples of the comparison between the correct $b$-dependence
of the solution and the anzatz  (\eq{Nb}).
It can be seen from the figure  Fig.~\ref{bdep}  that our anzatz  (\ref{Nb}) is a 
very good approximation
for relatively small $b$ and up to $b\sim R_A$. For larger impact parameters our anzatz 
underestimates the true solution. This fact can be simply understood theoretically. At very
large impact parameters   the exponential in  (\ref{Nb}) can be expanded and we get
$N_A = \kappa_A  S(b)/S(0)$. Then  substitution
of the anzatz to the equation (\ref{EQ}) shows that it underestimates the solution 
indeed. For such values of the impact parameter it is more natural to search for the true solution
of  (\ref{EQ}) in the form  $N_A(r_\perp,x; b)\,=\,n_A(r_\perp,x)\,S(b)/S(0)$ and to write down an
evolution equation directly for the function $n_A$. In this case, we would obtain 
 the function $n_A$ somewhat  larger than our $\kappa_A$.
Such an approach was adopted in the Ref. \cite{KKM}
but for the whole range in $b$ which seems to be wrong for small impact parameters and
the Glauber initial conditions (\ref{ini}).

\section{ Gluon density and the $F_2$ structure function}

\subsection{Gluon density  $xG_A$}

\begin{figure}[htbp]
\begin{tabular}{c c c}
 $xG_A/A$ &$xG_A/A $ 
& $xG_A/A$ \\
 \epsfig{file=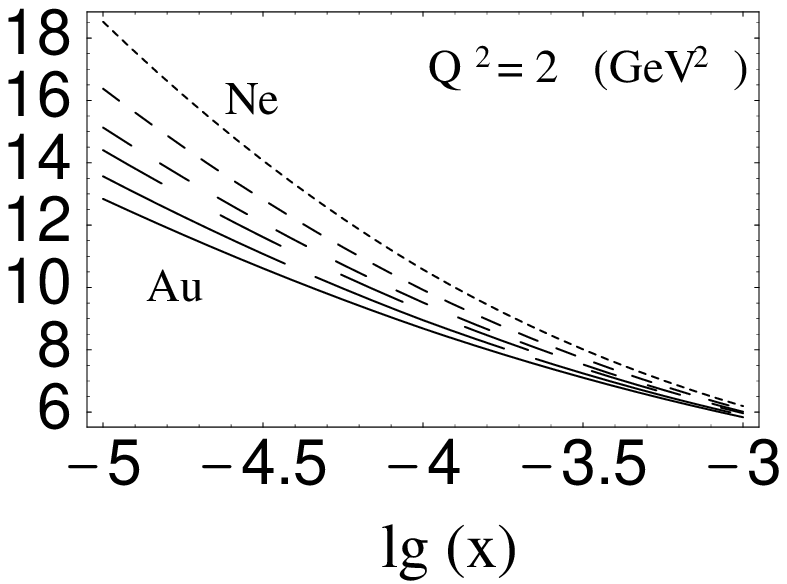,width=50mm, height=50mm}&
\epsfig{file=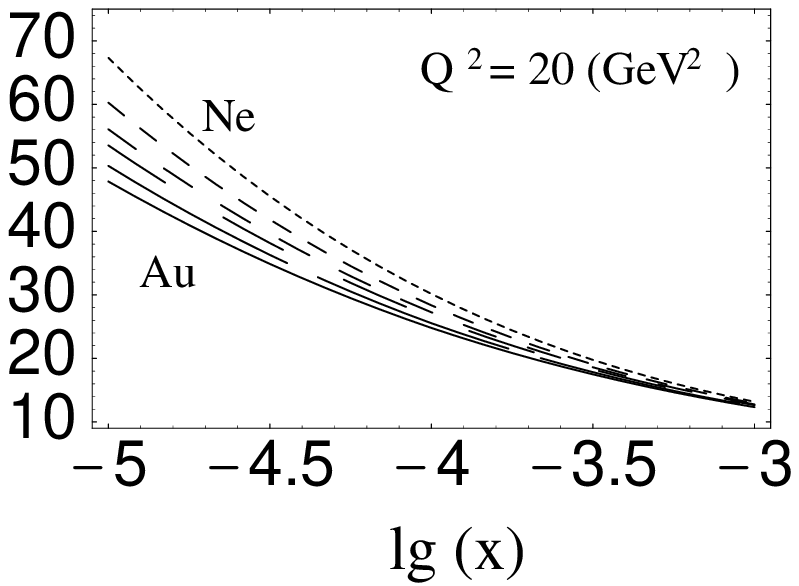,width=50mm, height=50mm} &
 \epsfig{file=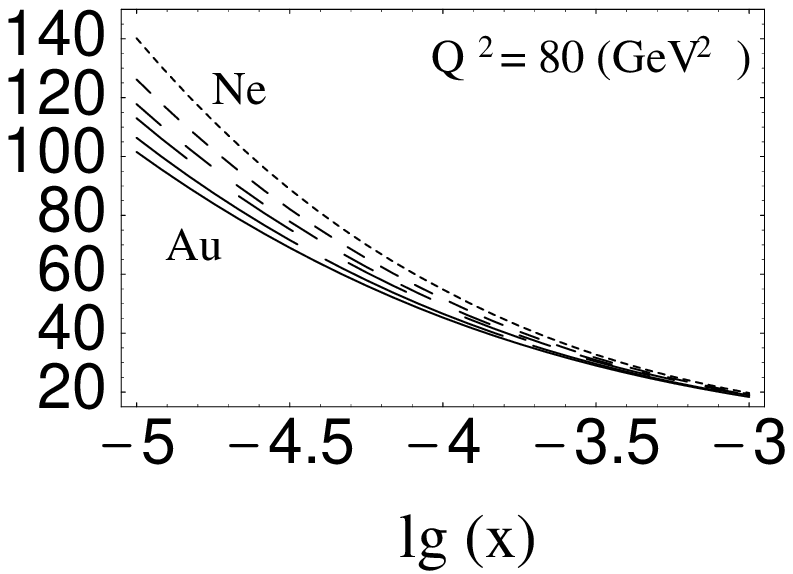,width=50mm, height=50mm}\\ 
 $xG_A/A^{2/3}$  & $xG_A/A^{2/3}$
 & $xG_A/A^{2/3} $ \\
 \epsfig{file=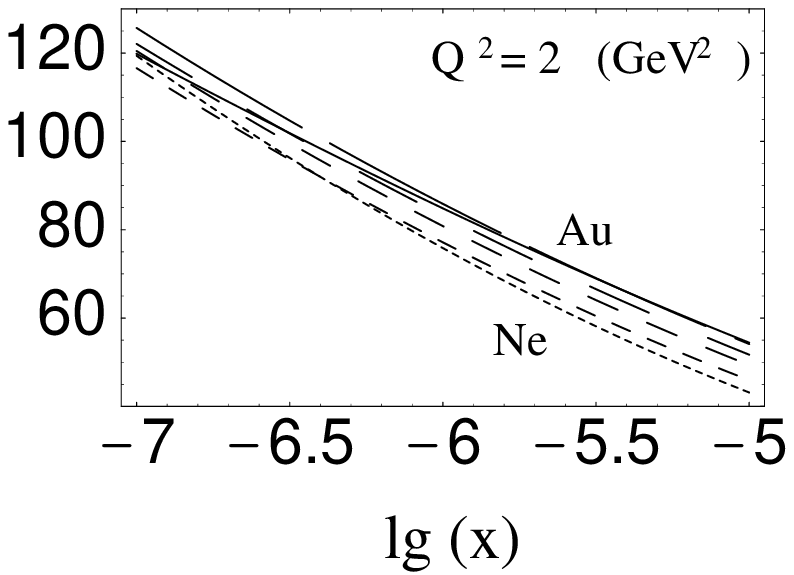,width=50mm, height=50mm}&
\epsfig{file=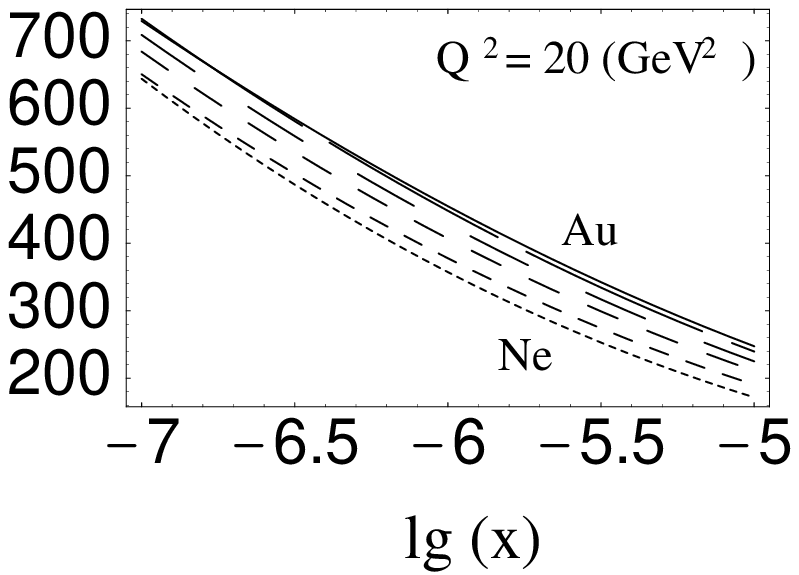,width=50mm, height=50mm}&
\epsfig{file=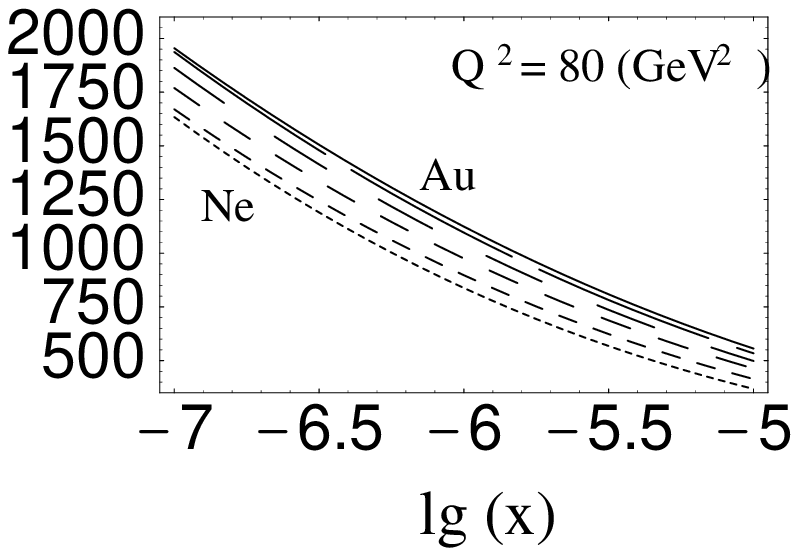,width=50mm, height=50mm}\\ 
\end{tabular}
\caption[]{\it The gluon density $xG_A$ is plotted as a function of $\lg x$ for the
nuclei  $Ne$, $Ca$, $Zn$, $Mo$, $Nd$, and $Au$.  }
\label{xGAplot}
\end{figure}

\begin{figure}[htbp]
\begin{tabular}{c c c}
 \epsfig{file=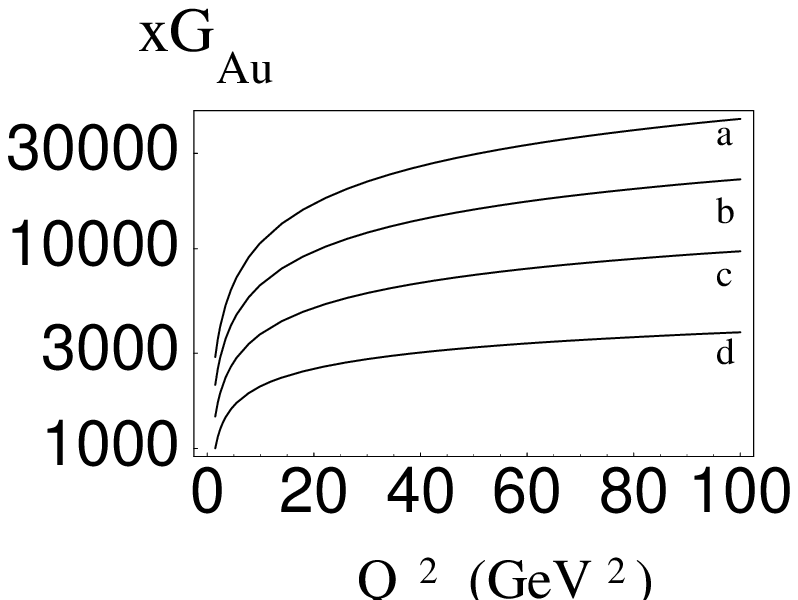,width=52mm, height=55mm}&
\epsfig{file=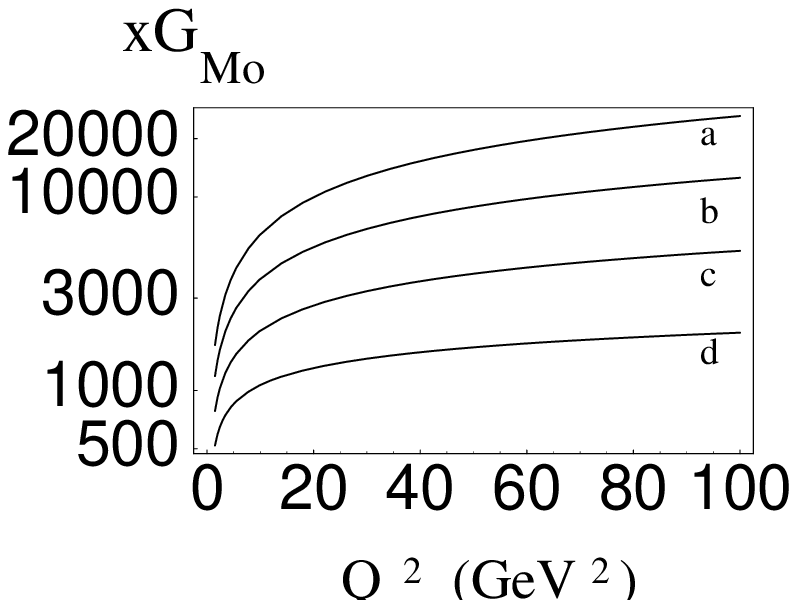,width=52mm, height=55mm}&
\epsfig{file=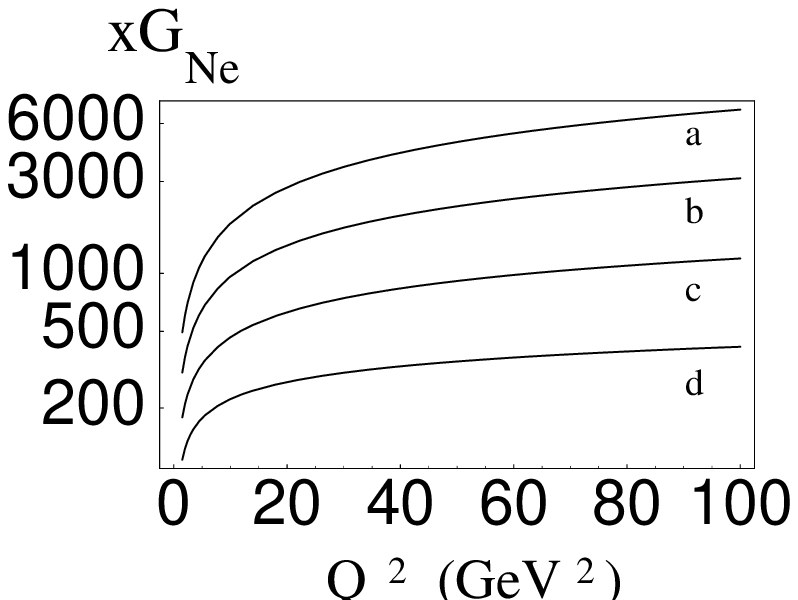,width=52mm, height=55mm}\\  
\end{tabular}
\caption[]{\it The   gluon density $xG_A$ is
plotted as a function of $Q^2$; a - $x=10^{-6}$; b - $x=10^{-5}$; c - $x=10^{-4}$;
 d - $x=10^{-3}$.}
\label{xLogGA}
\end{figure}

We can proceed now with the computation of the gluon density $xG_A$ which we define
using the Mueller formula \cite{DOF3}:
\beq
\label{xG}
xG_A(x,  Q^2)\,=\,\frac{4}{\pi^3} \int_x^1 \frac{dx^\prime}{x^\prime}
\int_{4/Q^2}^\infty\frac{dr^2}{r^4} \,\int d^2b \,2\,N_A(r,x^\prime;b)\,.
\eeq
There is some  uncertainty in (\ref{xG})   due to $b$-integration. Supposing that the main
contributions to the $b$-integration come from the region $b\sim R_A$, the uncertainty
can be estimated. In this region the anzatz  (\ref{Nb}) underestimates the solution by 
maximum about 25\%, 
and hence the  obtained gluon density is also expected to be underestimated 
by the same amount.  
The Fig. \ref{xGAplot} presents results of the calculations for the
gluon density $xG_A$ for three values of $Q^2$ while the Fig. \ref{xLogGA} shows its
$Q^2$ dependence  for various $x$.

The gluon density  $xG_A$ defined as in (\ref{xG}) 
grows with decreasing $x$. In a sense,
 this   observation contradicts the ``super-saturation'' of Ref. \cite{Braun} where the density is 
predicted to vanish in the large rapidity limit. However, the gluon density definition accepted
in Ref. \cite{Braun} differs from ours, and hence the functions are not compatible.

It is important to investigate the $A$ dependence of the gluon density $xG_A$. The leading
twist perturbative QCD prediction based on DGLAP equation is $xG_A=A\,xG_N$, where
$xG_N$ is a nucleon gluon density. 

Let parameterize the  $A$ dependence as $xG_A=A^n\,xG_N$. The results obtained 
for the power $n$ are presented in the table~(\ref{txGA}).
 At high $x$ we indeed obtain the law $xG_A\simeq A\,xG_N$
which is shown in the Fig.~\ref{xGAplot}. At moderate $x\simeq 10^{-5}$  a transition 
occurs and $n$ turns out to be quite different from 1. At low $x$ the gluon density $xG_A$ is rather
 proportional to $A^{2/3}$  
(Fig.~\ref{xGAplot}).  For large $Q^2$ the transition occurs
at somewhat smaller $x$, but the dependence on $Q^2$ is very weak   (Table ~\ref{txGA}).
This described $A$ dependence is actually  anticipated from general theoretical 
arguments (see Ref. \cite{AGL} and references therein).
\begin{table}
\begin{minipage}{9.0 cm}
\center{
\begin{tabular}{||l||c|c|c|c|c||} 
$Q^2\,\,\backslash \,\,x$ & $10^{-7}$ & $10^{-6}$ & $10^{-5}$ & $10^{-4}$ & $10^{-3}$  \\
\hline \hline
$ 2 \,GeV^2$&                                       0.73    &           0.78  &    0.84   &     0.92     & 0.97        \\  
\hline
$ 15 \,GeV^2$&                                    0.74     &       0.79   &    0.85      &     0.92    &   0.97     \\
\hline
 $40 \,GeV^2$&                                    0.75     &       0.8   &    0.86      &     0.92    &   0.97     \\
\hline
$ 120  \,GeV^2$ &                                        0.76     &       0.81   &    0.87     &     0.92    &   0.97 \\
\hline
\end{tabular}}
\end{minipage}
\begin{minipage}{7.0 cm}
\caption{The power $n$ for various values of $x$.}
\label{txGA}
\end{minipage}
\end{table}

\subsection{ Anomalous dimension}

We define the average anomalous dimension as follows   \cite{AGL}:

\beq
\label{gamma}
\gamma (A,x,Q^2)=\frac{d (\ln xG_A(x,Q^2))}{d \ln Q^2}\,.
\eeq

\begin{figure}[htbp]
\begin{tabular}{c c c}
(a) & (b) & (c) \\
 \epsfig{file=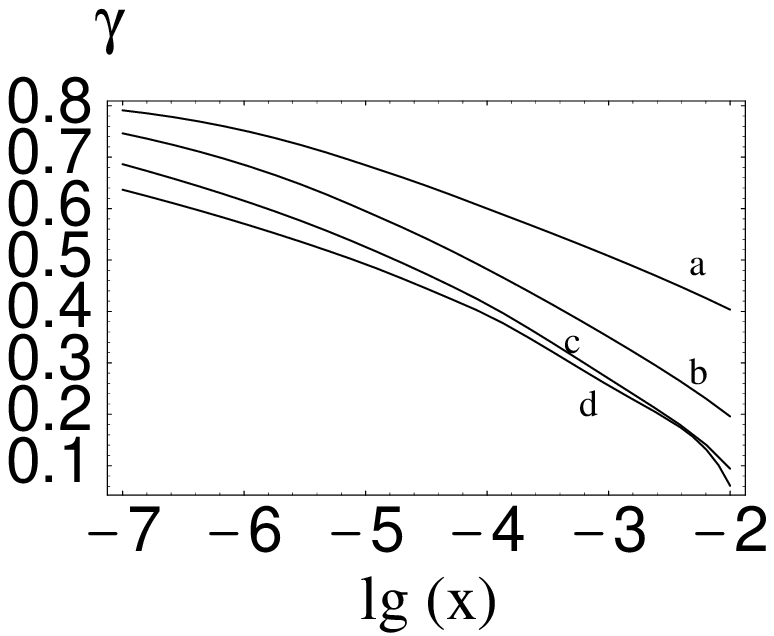,width=50mm, height=50mm} &
\epsfig{file=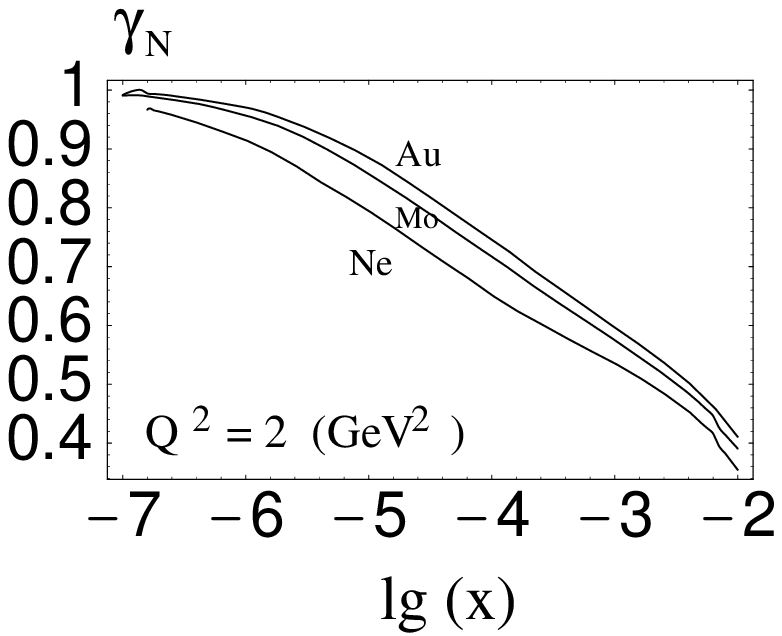,width=50mm, height=50mm} &
\epsfig{file=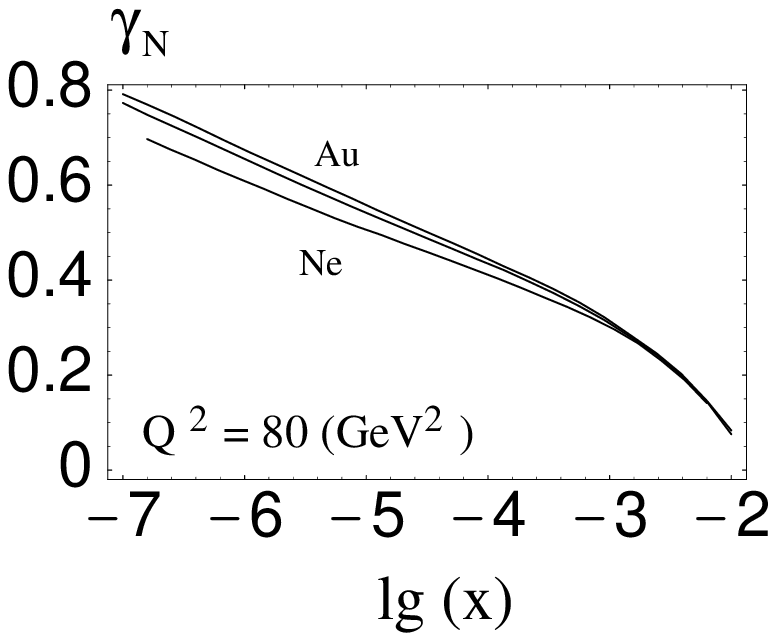,width=50mm, height=50mm}\\
\end{tabular}
\caption[]{\it The anomalous dimensions are
plotted as a function of $\lg x$. (a) -  $\gamma$ for the various values of 
$Q^2$: a - $1.5\,GeV^2$, b - $5\,GeV^2$, c - $30\,GeV^2$, d - $120\,GeV^2$; 
(b) - $\gamma_N$
for $Q^2=2\,GeV^2$; (c)  - $\gamma_N$
for $Q^2=80\,GeV^2$.}
\label{gammaplot}
\end{figure}

The results of the computations are presented on the Fig.~\ref{gammaplot}-a. To our surprise
the anomalous dimension almost does not 
depend on the atomic number $A$ displaying independence
on the target.  However, the anomalous dimension for the function $\tilde N_A$
satisfying \eq{EQ} shows a considerable $A$ dependence as can be seen in
the Fig.~\ref{gammaplot} (b-c).  We define $$
\gamma_N \,\,=\,\,\frac{d \ln \tilde N_A(2/Q,x)}{ d \ln Q^2}\,\,+\,\,1\,\,.$$
Such a different behavior of $\gamma$ and $\gamma_N$, in our opinion, is mostly
related to the fact that shorter distances enter the calculation of $\gamma$
compared to $\gamma_N$ as  can be observed directly from \eq{gamma}.

\subsection{$F_{2A}$}

The structure function $F_{2A}$  is related to the dipole cross section

\beq
\label{F_2}
F_{2A}(x,  Q^2)\,=\,\frac{2}{\pi^3} 
\int_{4/Q^2}^\infty\frac{dr^2}{r^4} \,\int d^2b \,N_A(r,x;b)\,.
\eeq

\begin{figure}[htbp]
\begin{tabular}{c c c}
 $F_{2A}/A $  &  $F_{2A}/A $ &  $F_{2A}/A $ \\
 \epsfig{file=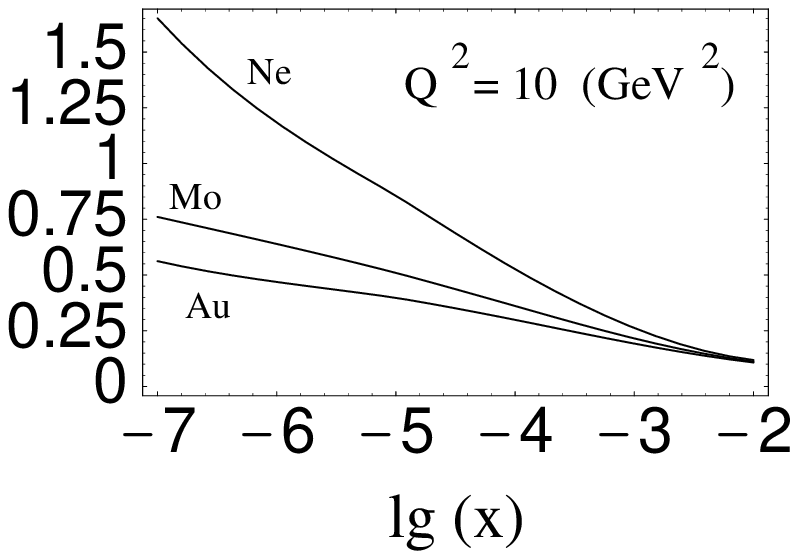,width=50mm, height=50mm}&
\epsfig{file=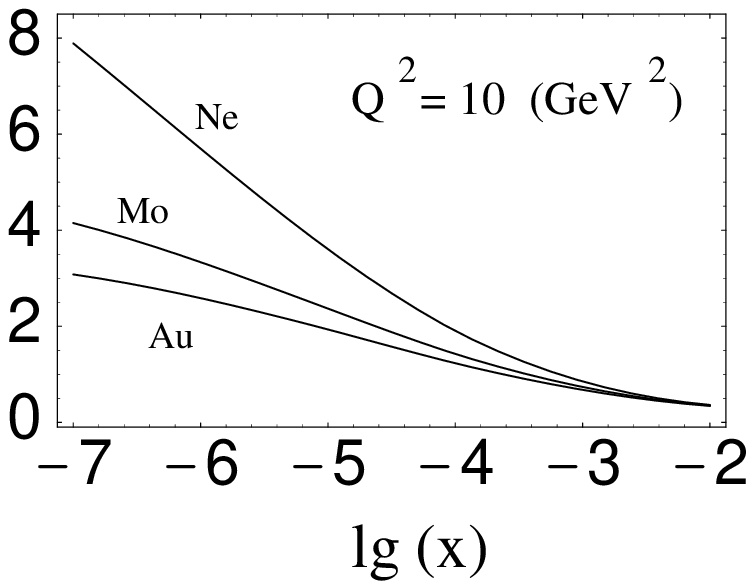,width=50mm, height=50mm}&
\epsfig{file=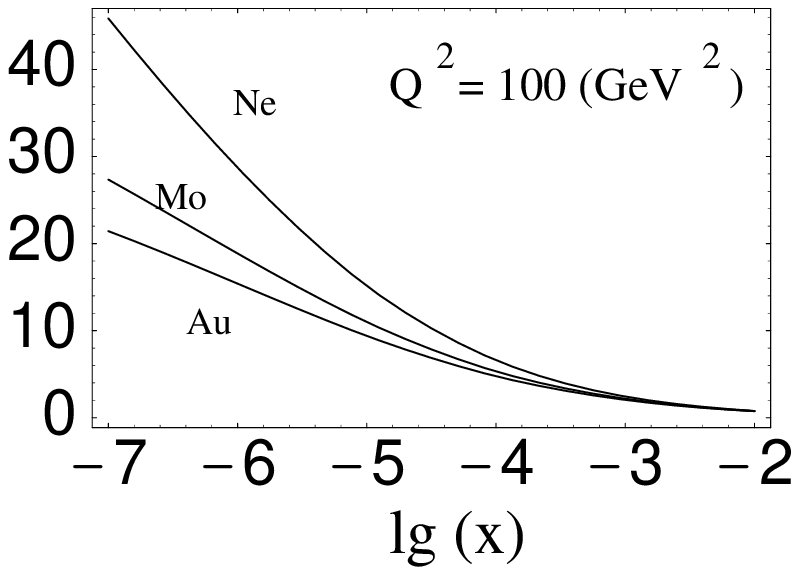,width=50mm, height=50mm}\\  
\end{tabular}
\caption[]{\it The  structure function $F_{2A}/A$ is
plotted as a function of $\lg x$ for the nuclei $Au$, $Mo$, and $Ne$.}
\label{F2plot}
\end{figure}

The Fig. \ref{F2plot} presents the function $F_{2A}$ for various nuclei.

We now in the position when we can try to compare our results with ones obtained 
in the Ref. \cite{Braun}. That paper presents results for the structure function $F_2$
computed for  the lead nuclei (A = 207).
 Unfortunately, the kinematic regions investigated 
in our work and in the Ref.  \cite{Braun} have  a very small overlap. Nevertheless at not
extremely  low $ x$ and not too high $Q^2\le 1000\,GeV^2$ a comparison can be made
(we use our gold ($A=197$) calculations). We fail to reproduce results of the Ref.  \cite{Braun}.
In all the region of the comparison we predict a few times smaller values for the function
$F_{2A}$.

There are several reasons for the obtained mismatch. First of all, there are  few minor ones 
related to the  difference between the  nuclei compared and different $\alpha_s$ used in 
the computations.
In addition, our treatments of the $b$-dependence are quite different.
However, in our opinion, the main reason is in the initial conditions of the evolution, which
do not coincide with the ones of  the Ref. \cite{Braun}. At moderate
 $x$ ($ x  = 10^{-3} \div 10^{-5}$)  our solution depends on initial condition as well as
 at lower $x$ where such a dependence is concentrated at very short distances.

Concluding this comparison we would like to make the following comment. We actually believe
that our results on $F_{2A}$ are more reasonable. At not very low $x\simeq 10^{-2}\div 10^{-4}$ 
we expect the function $F_{2A}$ to be proportional to $A$. Namely, 
$F_{2A}\simeq A\, F_{2}$. Our results on $F_{2A}$  display this behavior and are 
in a good agreement with the experimental data on the proton structure function $F_2$. 
Contrary to us, the results of  \cite{Braun} seem to be few times  off from this agreement.

\section{Saturation}

It is natural to define the saturation scale through the equation
\beq \label{satscale}
 \kappa_A(x,2/Q_s)\,=\,1/2.
\eeq
The definition (\ref{satscale}) agrees 
with the one adopted for the Glauber formula (\ref{GLAUBSI}).

\begin{figure}[htbp]
\begin{tabular}{c c}
\epsfig{file=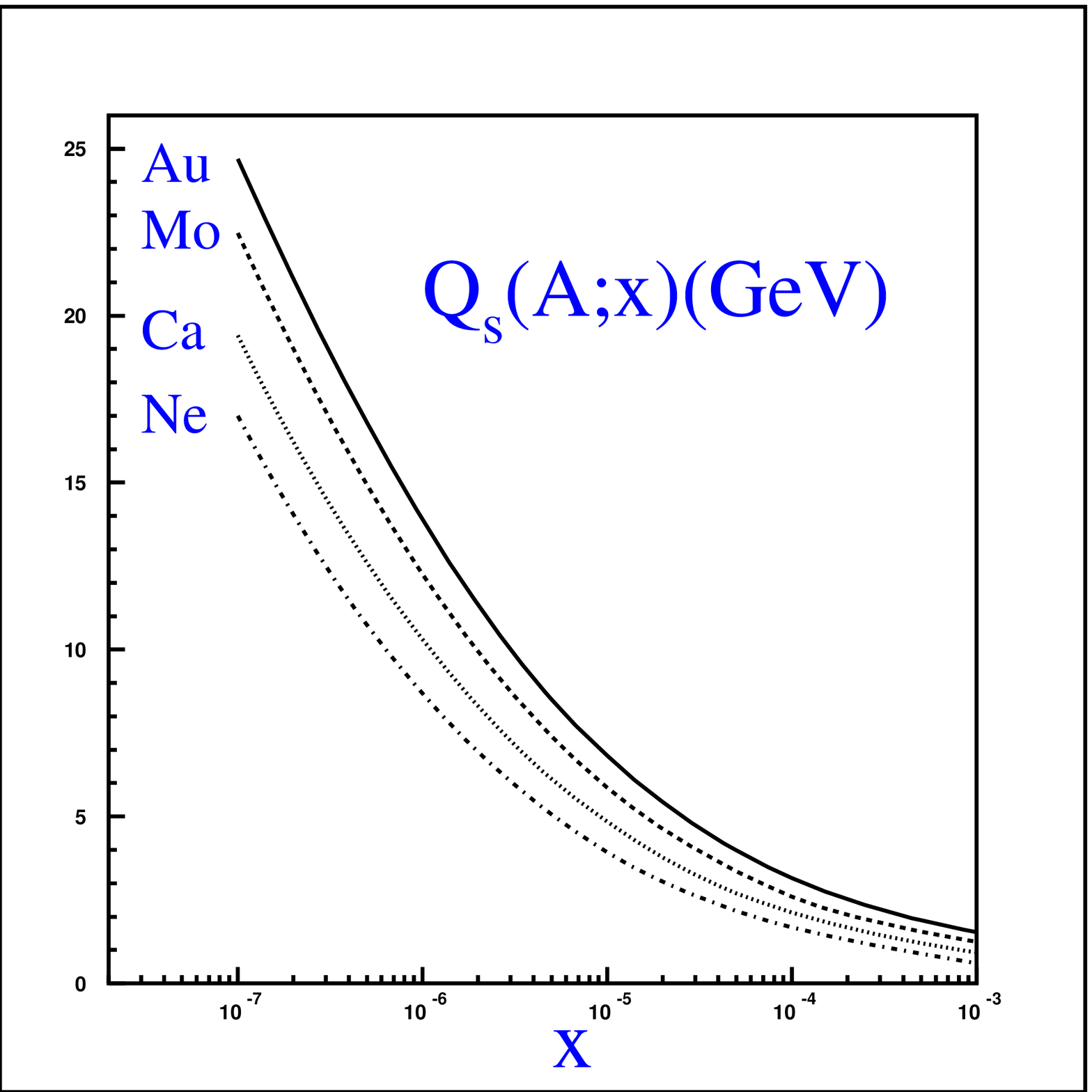,width=80mm, height=70mm} &
\epsfig{file=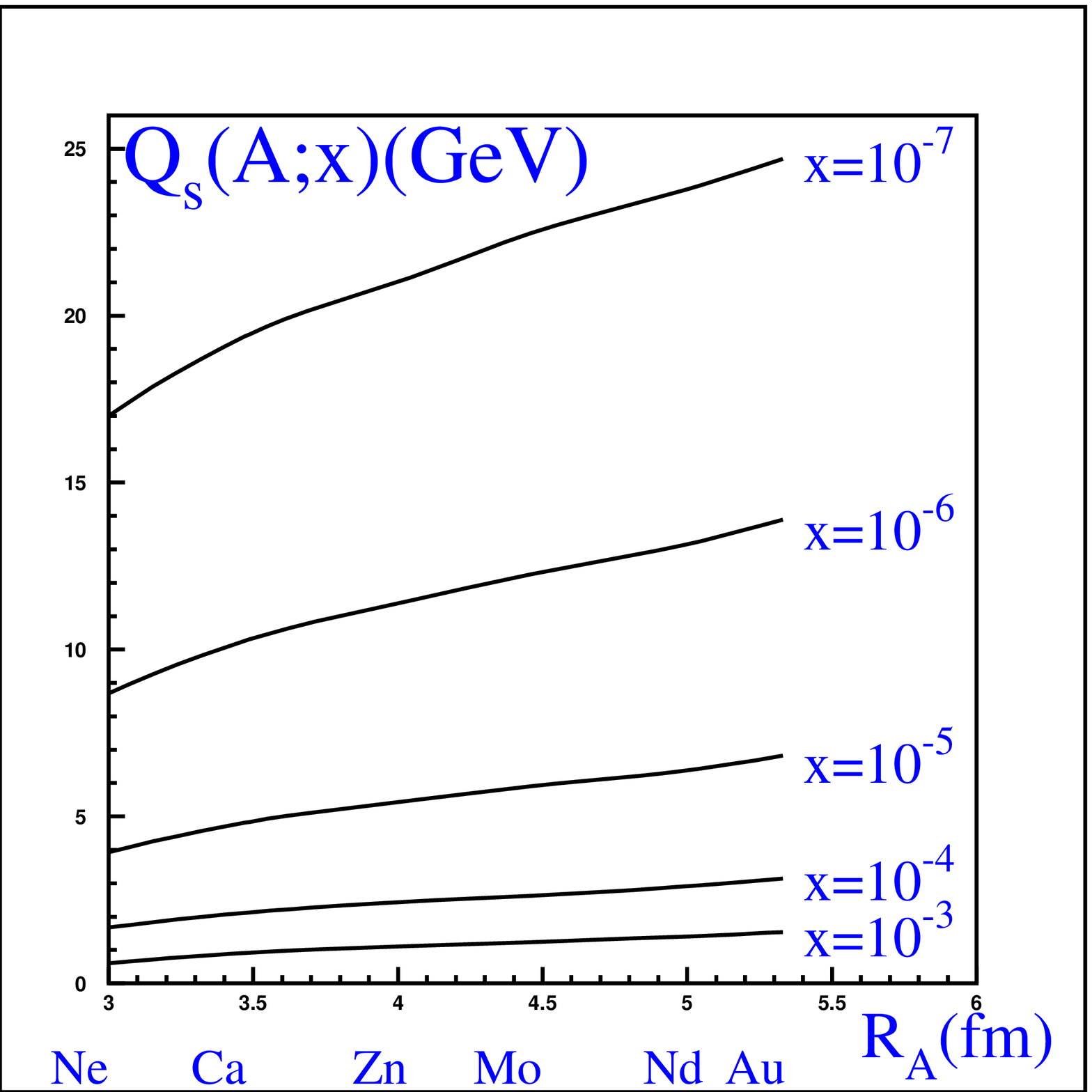,width=80mm, height=70mm}\\
(a) & (b)
 \end{tabular}
 \caption[]{\it The saturation scale  $Q_{s,A}$  is plotted versus  $\lg x$ (a) and $R_A$ (b).  }
\label{scale}
\end{figure}

Fig.~\ref{scale} displays the saturation scale $Q_{s,A}$ obtained in (\ref{satscale}) for various
nuclei.

The main question which we want to study is the dependence of the saturation scale 
$Q_{s,A}(x)$
on the atomic number $A$. 
Let us assume a power law behavior for the saturation scale $Q_{s,A}(x)$ as a function of $A$:
\beq
\label{power}
 Q_{s,A}(x)\,=\, C (x) \,A^{p(x)},
\eeq
where $C(x)$ and $p(x)$ are $x$ dependent functions. The power $p(x)$ is of our main
interest. In order to check the anzatz (\ref{power}) and   find the power 
 we study the $A$ dependence of the saturation scale  in  double logarithmic
scale  and find it to  fit well a straight line that
justifies use of the anzatz  (\ref{power}). The power $p(x)$  can be 
found by the least square fit. The table~(\ref{tA}) presents the fit for the three cases: light nuclei
($Ne$, $Ca$, $Zn$); heavy nuclei ($Zn$, $Mo$, $Nd$, and $Au$); all nuclei together.

\begin{table}
\begin{minipage}{9.0 cm}
\center{
\begin{tabular}{||l||c|c|c|c|c||} 
Nuclei $\,\,\backslash \,\,x$ & $10^{-7}$ & $10^{-6}$ & $10^{-5}$ & $10^{-4}$ & $10^{-3}$  \\
\hline \hline
 Light &                                       0.18    &           0.22  &    0.26   &     0.31     & 0.49        \\  
\hline
 Heavy &                                    0.15     &       0.19   &    0.22      &     0.24    &   0.31     \\
\hline
 All   &                                        0.17     &       0.20   &    0.24     &     0.27    &   0.39    \\
\hline
\end{tabular}}
\end{minipage}
\begin{minipage}{7.0 cm}
\caption{The power $p(x)$ for various values of $x$.}
\label{tA}
\end{minipage}
\end{table}
The table~(\ref{tA}) is quite transparent. The power $p(x)$ decreases with decreasing  $x$.
At the beginning
of the evolution the $A$-dependence of the saturation scale for the light nuclei 
is close to the law  $Q_{s,A}\sim A^{1/2}$, 
at  $x\simeq 10^{-4}$ the saturation scale
$Q_{s,A}\sim A^{1/3}$, while at higher energies it tends to $Q_{s,A}^2\sim A^{1/3}$.
For the heavy nuclei  the situation is similar but the decrease of the power $p(x)$ is significantly
slower. 
The above observations are in complete contradiction with conclusions derived in the
Ref. \cite{LT1}, where the saturation scales were deduced from the equation (\ref{EQ})
in the double logarithmic approximation. The main source of this large
discrepancy is the fact that the anomalous dimension in the solution of the
DGLAP equation turns out to be larger than 1/2 which is the maximum value for
the BFKL evolution in the leading order (see \eq{SATSCEST}). This is a
disturbing result since it could mean that the leading order non-linear
evolution equation (see \eq{EQ}) is not enough to make a reliable
predictions in the kinematic region of high density QCD.

We would like to stress that the numbers presented
in the table (\ref{tA}) are quite approximate. These numbers display a certain 
 sensitivity to the definition of the saturation scale (we use \eq{satscale} but other 
definitions can be investigated as well). Moreover, the powers obtained are results of 
regressions over too small number of points, which actually implies large errors. However,
we are convinced that the decreasing property of the power $p(x)$ with high energies 
is quite general. 

Table~(\ref{tA}) can be explained qualitatively in terms of the anomalous dimension $\gamma$. 
The following arguments are presented for the GM formula (\eq{GLAUBSI}) or/and 
for the packing factor (\eq{KAPPA}), but the physical
picture is valid for the solution of the equation (\ref{EQ}) as well. 
In the GM formula the power $p(x)=\frac{1}{6(1-\gamma)}$. Though $\gamma$ is usually
assumed to be small, it is often not the case. Moreover,  $\gamma$ is actually $x$ and
$Q$ dependent. At large $x$ and  correspondingly small saturation scale  
$Q_{s,A}$ the GRV gluon density parameterization  implies very large $\gamma$ tending
to unity (see e.g. Ref. \cite{AGL}). This is the origin of the large powers
in the table (\ref{tA}). When we go from light nuclei to more  heavy (at  fixed $x$) the saturation
scale obtained increases, 
and consequently $\gamma$ decreases. That is why smaller powers are 
obtained for the heavy nuclei compared to the light ones. As $x$ decreases, on one hand
 $\gamma$ increases at fixed $Q$. On the other hand, at smaller $x$ a larger  saturation scale
 $Q_{s,A}$ is obtained. Finally, with decreasing $x$ the power $p(x)$ decreases as well tending
to the value $p(x)=\frac{1}{6}$. 

\section{Discussion of the results}

Let us explain qualitatively the $ A$ dependence of the results. We also wish to perform
checks for the self-consistency of the results obtained.

The main starting 
observation is that the anomalous dimension $\gamma$ (\ref{gamma})
is almost $A$ independent while its 
$Q^2$ dependence become   weaker as $x$ decreases.

Define the power $\alpha$:

\beq
\label{alpha}
\alpha (A,x,Q^2)\,=\,-\,\frac{d (\ln \kappa_A(x,2/Q))}{d \ln Q^2}\,.
\eeq

We find  that $\alpha  \simeq 1-\gamma $.
Consequently we can try to write $\kappa_A$ in the power-like form:

\beq
\label{kapp}
\kappa_A(x,r_\perp)\,\sim\, A^{\beta(x)}\, (r_\perp^2)^{\alpha} \,.
\eeq

In the equation (\ref{kapp}) we introduced the $A$-dependence in order to define 
the saturation scale:

\beq
Q_{s,A}(x)\, \sim \,A^{\beta/(2\alpha)}\, =\,A^{p(x)}\,.
\label{sat}
\eeq

The power $\beta$ is obtained to be almost $Q^2$ independent. It ranges from
about 0.33 at $x=10^{-3}$ and to about 0.17 at $x=10^{-7}$. Meanwhile at the saturation
scale $\alpha$ is almost constant $\alpha\simeq 0.45\div 0.5$. This corresponds
to $\gamma \simeq 0.5$, which is the anomalous  dimension of the linear BFKL term
in the nonlinear equation (\ref{EQ}).

Assuming  $\kappa_A$  be relatively small we would obtain for the gluon density

\beq
\label{gluon}
xG_A(x,Q^2)\,\sim\, R_A^2\, \int_x^1 \frac{dx^\prime}{x^\prime}\int_{4/Q^2} \frac{dr^2}{r^4}\, 
\kappa_A(x^\prime,r)\,.
\eeq

Substituting (\ref{kapp}) to the equation (\ref{gluon}) we obtain

\beq
\label{gluon1}
xG_A(x,Q^2)\,\sim\, R_A^2 \,\int_x^1 \frac{dx^\prime}{x^\prime}\, (Q^2)^{1-\alpha}\, 
A^\beta\,.
\eeq
Assuming that the integral in (\ref{gluon1}) is dominated by $x^\prime\simeq x$,
we see  that within the above approximation  $\gamma\,simeq\,1-\alpha$ indeed. The 
independence of the anomalous dimension on $A$ is a consequence of the fact that
$\beta$ happened to be almost $Q^2$ independent.

What about $A$-dependence of the gluon density? We obtain
for the power $n$ the relation
\beq
\label{n}
n(x)\,= \,2/3\,+\,\beta(x)\, = \,2/3\, +\, 2\, p(x)\, \alpha(x)\,.
\eeq

The equation (\ref{n}) is in a sense an additional consistency check. The numbers presented
in the tables (\ref{txGA}) and (\ref{tA}) are in a good agreement with each other (though
the equation  (\ref{n}) is not fulfilled exactly due to many approximations done on the way
to (\ref{n})). The fact that the powers $n$ in the table  (\ref{txGA}) do not much 
sensitive to $Q^2$ is now understood as a consequence of the corresponding 
independence of $\beta$.

\section{Conclusions}

In this paper we reported on the exact numerical solutions 
of the nonlinear evolution equation (\ref{EQ}) for  nuclear targets. The solutions are
obtained by the method of iterations proposed in the Ref.  \cite{LGLM}. Various nuclei
starting from the very  light $ Ne_{20}$ and up to heavy $ Au_{197}$ were
investigated.

We demonstrated that the solution to the non-linear equation is quite 
different from the GM model  that has been used
for estimates of the saturation effects. 
However, this model can be used as a first
 iteration of the non-linear equation which leads to faster convergence of 
the numerical procedure.

From the experimental point of view the obtained results  support  the energetic profit
 for performing DIS experiments on nuclei. However, the gain is quite modest, and it can be
estimated about five times for gold targets compared with   proton. 

The gluon density $xG_A$ and the structure function $F_{2A}$ were estimated. 
At small $x\simeq 10^{-6}\div 10^{-7} $ the damping due to non-linear
effects leads to suppression of a factor  $2 \div 4$ for heavy nuclei
 in comparison with the  DGLAP prediction $xG_A=A\,xG^{DGLAP}$. At moderate 
$x\simeq 10^{-2} \div 10^{-4} $ the structure function $F_{2A}\simeq A\, F_2$, where the 
obtained  values for $F_2$ agree well with the experimental data on proton.

The dependence of the gluon density $xG_A$ was investigated as a function of the atomic
number $A$. At high $x$, the density is shown to be proportional to $A$.
A transition occurs at moderate $x\simeq 10^{-4}$. 
At small $x$, $xG_A$ is rather proportional to $A^{2/3}$.  To our knowledge this is a first 
numerical confirmation  of this dependence from the master equation \eq{EQ}  
expected on general grounds.

The found solutions of the nonlinear equation 
were used  to estimate the saturation scale
$Q_{s,A}(x)$.  In agreement with all theoretical predictions 
the  saturation scale grows with decreasing $x$. 

The dependence of the saturation scales on the atomic number $A$ was  a focus of 
the research. A fit of the saturation scale to the power law $Q_{s,A}(x)\sim A^{p(x)}$ was 
investigated. The numerical values obtained for the power $p(x)$ are actually  
sensitive to the saturation scale definition. Nevertheless, we predict a decreasing behavior
of the power with decreasing $x$ for both heavy and light nuclei.

One of the interesting properties of the solutions of the equation (\ref{EQ}) is that
they display the scaling phenomena. Namely, the solution $\tilde N_A$ is not a function
of two variables $x$ and $r_\perp$ but rather a function of a single variable 
$r_\perp\,Q_{s,A}(x)$. A complete report on this subject is now in a final stage of preparation
and will appear shortly \cite{me}.

Among several observations which come from the numerical solutions we would
like to point out that the weak dependence of  $xG_A$ anomalous dimension 
 on the atomic numbers as well as  the fact that $Q^2_{s,A} \propto A^{1/3}$ at $x
\rightarrow 0$ look quite impressive  for us. They certainly need some
theoretical explanation and we hope they will stimulate such
explanations. We firmly believe that our calculations of $Q_{s,A}$ for various
nuclei  will provide a theoretical basis for the 
discussion of the RHIC data in high parton density QCD approach \cite{RHIC}.

When the present paper was finished for publication we read  a new paper 
\cite{Braun2} devoted to the very same subject. We have to stress that
 the initial conditions of  our analyses are quite different. Though within the evolution in
$x$ the influence of   initial conditions become weaker  they are important for quantitative
analysis in the experimentally reasonable region of not too small $x$. We were pleased 
to discover that despite the difference 
in the initial conditions the saturation scale $Q_s$ obtained in \cite{Braun2} is proportional
to $A^{2/9}$, which is a dependence 
 quite similar to ours.  The scaling phenomena  confirmed by the numerical calculations of 
the Ref. \cite{Braun2} has  been also observed by one of us and the report will appear soon
 \cite{me}.

{ \bf Acknowledgments :} 

We are very grateful to Ia. Balitsky, M. Braun, E. Gotsman, E. Iancu,
D. Kharzeev, Yu. Kovchegov,  U. Maor,  L.   McLerran 
  and K. Tuchin 
   for illuminating discussions  on the subject.

 This research was supported in part by the BSF grant $\#$
9800276, by the GIF grant $\#$ I-620-22.14/1999
  and by
Israeli Science Foundation, founded by the Israeli Academy of Science
and Humanities.

\end{document}